\documentstyle[epsfig,12pt]{article}  
\newcommand{\bce}{\begin{center}}
\newcommand{\ece}{\end{center}}
\newcommand{\beq}{\begin{equation}}
\newcommand{\eeq}{\end{equation}}
\newcommand{\bea}{\vspace{0.25cm}\begin{eqnarray}}
\newcommand{\eea}{\end{eqnarray}}

\newcommand{\ba}{\begin{array}}
\newcommand{\ea}{\end{array}}

\newcommand{\r}{\mbox{{\boldmath
$\rho$}}}

\newcommand{\bteta}{\mbox{\boldmath
$\theta$}}
\newcommand{\bvarphi}{\mbox{\boldmath
$\varphi$}}

\newcommand{\vb}{\mbox{{\bf
v}}}
\newcommand{\wb}{\mbox{{\bf
w}}}

\newcommand{\rb}{\mbox{{\bf
r}}}

\newcommand{\kb}{\mbox{{\bf
k}}}
\newcommand{\xb}{\mbox{{\bf
x}}}

\def\lsim{\mathrel{\rlap{\lower4pt\hbox{\hskip1pt$\sim$}}
    \raise1pt\hbox{$<$}}}         
\def\gsim{\mathrel{\rlap{\lower4pt\hbox{\hskip1pt$\sim$}}
    \raise1pt\hbox{$>$}}}         

    \def\beq{\begin{equation}}
    \def\endeq{\end{equation}}
    \def\bea{\begin{eqnarray}}
    \def\arr{\begin{eqnarray}}
    \def\eea{\end{eqnarray}}

\def\q2{$Q^{2}$}
\def\s2{2$S$}

\begin{document}
\thispagestyle{empty}
\vspace*{-2cm}
\begin{flushright}
{\bf\large
FZJ-IKP(Th)-1999-22\\}
\end{flushright}
 
\bigskip

\begin{center}

  {\large\bf
COMMENT ON THE BAIER-KATKOV QUASICLASSICAL OPERATOR APPROACH 
TO THE LANDAU-POMERANCHUK-MIGDAL EFFECT
\\
\vspace{1.5cm}
  }
\medskip

{\bf \large 
B.G. Zakharov
\bigskip\\}
{\sl 

Institut  f\"ur Kernphysik, Forschungszentrum J\"ulich,\\
        D-52425 J\"ulich, Germany\medskip\\
Landau Institute for Theoretical Physics,
        GSP-1, 117940,\\ Kosygina Str. 2, 117334 Moscow, Russia
}
\vspace{4cm}

  {\bf Abstract}
\end{center}
I demonstrate that the Baier-Katkov approach to the Landau-Pomeranchuk-Migdal
effect based on their quasiclassical operator method is  
conceptually wrong in quantum regime of interaction of the charged particles
with medium constituents which takes place for 
electrons and positrons in real media.

\newpage

Ever since the celebrated  papers by Landau and Pomeranchuk \cite{LP}, 
and Migdal \cite{Migdal} the influence of multiple scattering on the
radiation processes in medium (the so-called Landau-Pomeranchuk-Migdal
(LPM) effect) attracted much attention
(see, for instance, the monograph \cite{TM} and recent review \cite{Klein}).
The new activity in the LPM effect in QED  has been stimulated  by the first 
accurate experimental investigation of this effect by the SLAC 
E-146 collaboration \cite{SLAC}.
In Ref. \cite{BGZ1} (see also Refs. \cite{BGZ2,BGZ3,BGZ4,BGZ5}) 
I have developed a new rigorous light-cone path integral (LCPI) 
approach to the 
LPM effect in QED and QCD. I have expressed the radiation rate through
solution of a two-dimensional Schr\"odinger equation with an 
imaginary potential. 
The representation for the radiation rate derived in Ref. \cite{BGZ1}
is similar to the one obtained long ago in Refs. \cite{BKF,BKS}
within the Baier-Katkov quasiclassical operator (QO) approach
to the radiation processes in QED \cite{BKQO}. 
In Ref. \cite{BKS} the radiation rate has been evaluated
for a medium without an external field, and in Ref. \cite{BKS}
for a medium with allowance for a smooth external field.
The representation of Refs. 
\cite{BKF,BKS} was recently used for analysis of the LPM effect
in several papers by Baier and Katkov \cite{BK1,BK2,BK3,BK4}.
Due to the coincidence of the final expressions for the radiation rate
of Ref. \cite{BGZ1} and Refs. \cite{BKF,BKS}, Baier and Katkov in 
Ref. \cite{BK1} conclude that the results of 
the LCPI approach \cite{BGZ1} to the LPM effect coincide with 
that of the QO method.

In this comment I would like to point out that 
the analysis of the LPM effect of Refs. \cite{BKF,BKS} is conceptually
wrong in the case of real media, and the approach developed cannot 
be regarded as a consistent
quantum theory of the LPM effect. As far as the coincidence
of the formulas for the radiation rate obtained in Refs. 
\cite{BKF,BKS} and in Ref. \cite{BGZ1} is concerned, it is of accidental 
nature. 
It is an artifact of an interplay of two conceptual errors
made by the authors of Refs. \cite{BKF,BKS}.
As will be seen below the incorrectness of the analysis of the LPM
effect given in Refs. \cite{BKF,BKS} is almost evident. Nonetheless,
it seems, this fact is not known among the experts.

Let us discuss for definiteness the 
$e\rightarrow e\gamma$ transition 
in an infinite medium in the presence of an external field considered 
in Ref. \cite{BKS}.  
Its authors assume that the emission probability 
can be evaluated by averaging over the classical electron
trajectories and atomic positions of the expression for 
the radiation rate given by the QO method for a given classical trajectory.
The corresponding QO formula (Eq.(2.1) of Ref. \cite{BKS}) reads
\beq
dw=\frac{\alpha}{(2\pi)^{2}}\frac{d^{3}k}{\omega}
\int dt_{1}\int dt_{2} R^{*}(t_{1})R(t_{2})
\exp\left\{-\frac{i\varepsilon}{\varepsilon'}[kx(t_{2})-kx(t_{1})]\right\}\,,
\label{eq:1}
\eeq
where $\alpha=1/137$, $k=(\omega,\kb)$ is the four-momentum of the photon,
$\varepsilon$ and $\varepsilon'$ are the initial and final electron energy,
$x(t)=(t,\rb (t))$, $t$ is the time, and $\rb$ is the electron
location on a classical trajectory, the factor $R(t)$ can 
be expressed through the electron spinors, its specific form is not
important for us. In Ref. \cite{BKS} the averaging over the 
trajectories and the atomic
positions is performed with the aid of the distribution
function, averaged over atomic positions. This leads to the
following expression (Eq. (2.4) of Ref. \cite{BKS}) for the emission 
probability per unit time 
\bea
dW=\left\langle\frac{dw}{dt}\right\rangle=
\frac{\alpha}{(2\pi)^{2}}\frac{d^{3}k}{\omega}
\mbox{Re}\int\limits_{0}^{\infty} d\tau
\exp\left(-\frac{i\varepsilon}{\varepsilon'}\omega\tau\right)\nonumber\\
\times
\int d^{3}\vb d^{3}\vb'd^{3}\rb d^{3}\rb' 
{\cal{L}}(\bteta',\bteta)F_{i}(\rb,\vb,t)
F_{f}(\rb',\vb',\tau;\rb,\vb)
\exp\left\{i\frac{\varepsilon}{\varepsilon'}\kb(\rb'-\rb)\right\}\,.
\label{eq:2}
\eea
Here, ${\cal{L}}(\bteta',\bteta)=2R^{*}(t+\tau)R(t)$, the angle 
$\bteta\approx \vb_{\perp}$
($\vb_{\perp}$ is the component of of electron velocity perpendicular
to the vector $\kb/\omega$). The distribution function $F_{i}$ satisfies
the kinetic equation
\beq
\frac{\partial F_{i}(\rb,\vb,t)}{\partial t}+
\vb\frac{\partial F_{i}(\rb,\vb,t)}{\partial \rb}
+\wb\frac{\partial F_{i}(\rb,\vb,t)}{\partial \vb}=
n\int d^{3}\vb' \sigma(\vb,\vb')[F_{i}(\rb,\vb',t)-F_{i}(\rb,\vb,t)]\,,
\label{eq:3}
\eeq
where $n$ is number density of the medium, $\wb$ is the electron
acceleration in the external field, and $\sigma(\vb,\vb')$
is the electron scattering cross section on the medium constituent (atom). 
The distribution function $F_{f}$ satisfies a similar equation.
The normalization condition for $F_{i}$ and initial one for $F_{f}$ read 
\beq
\int d^{3}\rb d^{3}\vb F_{i}(\rb,\vb,t)=1\,,
\label{eq:4}
\eeq
\beq
F_{f}(\rb',\vb',0;\rb,\vb)=\delta(\rb-\rb')\delta(\vb-\vb')\,.
\label{eq:5}
\eeq
Using Eqs. (\ref{eq:2})-(\ref{eq:5}) after some algebra the authors 
of Ref. \cite{BKS}  obtain the following expression for the 
spectral distribution of emission probability per unit time
(Eq. (2.18) of Ref. \cite{BKS})
\beq
\frac{dW}{d\omega}=
\alpha\omega
\mbox{Re}\int\limits_{0}^{\infty} d\tau
\exp\left(-i\frac{a\tau}{2}\right)
\left[\frac{\omega^{2}}{\gamma^{2}\varepsilon'^{2}}
\varphi_{0}(0,\tau)-i\left(1+\frac{\varepsilon^{2}}{\varepsilon'^{2}}\right)
\nabla\bvarphi(0,\tau)\right]\,,
\label{eq:6}
\eeq
where $\varphi_{\mu}$ satisfies equation
\beq
\frac{\partial \varphi_{\mu}(\xb,\tau)}{\partial \tau}-
\frac{ib}{2}\Delta_{x}\varphi_{\mu}(\xb,\tau)+
i\wb\xb\varphi_{\mu}(\xb,\tau)=n\varphi_{\mu}(\xb,\tau)\int d^{2}\bteta 
[\exp(i\bteta \xb)-1]\sigma(\bteta)\,,
\label{eq:7}
\eeq
$$
\varphi_{0}(\xb,0)=\delta(\xb),\,\,\,\,\,\,\,\,\,\,
\bvarphi(\xb,0)=-i\nabla \delta(\xb)\,,
$$
$a=\omega m_{e}^{2}/\varepsilon\varepsilon'$, 
$b=\omega\varepsilon/\varepsilon'$, $\gamma=\varepsilon/m_{e}$, 
$\sigma(\bteta)$ is the electron 
scattering cross section written through the angular variable $\bteta$.
A similar derivation of the radiation rate in the absence of
external field ($\wb=0$) has been given in Ref. \cite{BKF}.
By changing the two-dimensional variable $\xb$ by the 
transverse coordinate $\r=\xb/\omega$ 
one can cast the result of Refs. \cite{BKF,BKS} in the form obtained
in my paper \cite{BGZ1} (there I have discussed 
the situation without external field, however, it can be
trivially included). 
In Ref. \cite{BGZ1} the analogue of the right hand side of 
Eq. (\ref{eq:7}) contains the dipole cross section $\sigma_{d}(\rho)$
of scattering of $e^{+}e^{-}$ pair on the medium constituent.
In order to represent Eq. (\ref{eq:7}) in the form of Ref. \cite{BGZ1}
one should use the relation 
\beq
\sigma_{d}(\rho)=
2\int d^{2}\bteta 
[1-\exp(i\bteta \varepsilon  \r)]\sigma(\bteta)
\label{eq:8}
\eeq
between the differential cross section $\sigma(\bteta)$ and 
dipole cross section $\sigma_{d}(\rho)$. It is crucial that 
to represent Eqs. (\ref{eq:6}), (\ref{eq:7}) in the form
obtained in Ref. \cite{BGZ1} one should use the {\it quantum} electron 
scattering cross section $\sigma(\bteta)$.

My criticism of the derivation of the radiation rate given in 
Refs. \cite{BKF,BKS} is based on the following two facts:\\
1. The QO expression (\ref{eq:1}) is valid if the motion of the 
electron in the total potential (which equals the sum of
the medium potential and external one) is quasiclassical.
In the quasiclassical limit the typical transverse scale
at which the quantum interference of the electron trajectories
are important becomes considerably smaller than the typical
scale of variation of the medium constituent potential.
As a result, in such a regime in scattering on a medium constituent 
there exists an approximate correspondence between the scattering 
angle and the 
electron impact parameter, and the concept of the classical trajectory
is justified. Eq. (\ref{eq:1}) derived neglecting the variation
of the field acting on the electron for the quantum fluctuations
of the electron trajectory is valid only in this quasiclassical limit.
In the quantum regime the above correspondence between the scattering
angle and the electron impact parameter in interaction with a medium
constituent is lost. In this case one must take into account
accurately the variation of the field acting on the electron 
in evaluating the radiation rate. Eq. (\ref{eq:1}) (obtained
for a smooth field) does not make
any sense in the quantum regime. It is well known
that for a screened Coulomb potential the quasiclassical situation
takes place at $Z\alpha\gg 1$. For real media, when $Z\alpha < 1$,
we have essentially quantum regime. For this reason, it is evident that Eqs. 
(\ref{eq:1}), (\ref{eq:2}) are not justified for real media. 
\footnote{The fact that for a 
purely Coulomb potential the
quantum scattering cross section coincides with the classical one, of course,
does not justify the use of Eqs. (\ref{eq:1}), (\ref{eq:2}). In addition,
for the screened potential the quantum and classical cross sections
differ.}
The use of Eqs. (\ref{eq:1}) and (\ref{eq:2}) which are not valid
in the quantum regime is the main conceptual error in the treatment 
of the LPM effect given in Refs. \cite{BKF,BKS}.\\
2. The authors of Refs. \cite{BKF,BKS} compensate
the incorrectness of Eq. (\ref{eq:1}) by another 
error in the procedure of averaging over the electron trajectories.
According to their logic the QO expression (\ref{eq:1}) 
must be averaged over all possible classical trajectories
(and the authors say this). 
It is evident that in this case the distribution functions 
entering (\ref{eq:2}) should satisfy
the kinetic equation in which the collisional term contains 
the {\it classical} scattering cross section. However, the authors
use in the kinetic equation (\ref{eq:3}) the quantum cross section.  
As was said only in this case the formula for the radiation rate of 
Refs. \cite{BKF,BKS} coincides with that of Ref. \cite{BGZ1}.

The above facts make it clear that the approach to the LPM effect of Refs. 
\cite{BKF,BKS} based on the Baier-Katkov QO method 
cannot be regarded as a consistent quantum theory 
of the LPM effect. It is only justified in the quasiclassical limit 
which does not take place for real media.
The formal coincidence of the final expression for the radiation 
rate obtained in Refs. \cite{BKF,BKS} with the one obtained 
by me in the LCPI approach \cite{BGZ1} is of accidental nature. 
It is a consequence 
of an interplay of the use by the authors of Refs. \cite{BKF,BKS} 
of the incorrect expression (\ref{eq:2}) and replacement of 
the classical scattering cross section by the quantum
one in the kinetic equation (\ref{eq:3}).

In conclusion I would like to emphasize that my criticism concerns 
namely the conceptual aspects of the Baier-Katkov QO approach to the 
LPM effect, and does not concern the technical details of 
the subsequent analyses by Baier and Katkov \cite{BK1,BK2,BK3,BK4}. 
However, one has to bear in mind that the starting expression 
for the radiation rate used in Refs. \cite{BK1,BK2,BK3,BK4} has never 
been rigorously derived within the Baier-Katkov QO method.
\\
\vspace{.5cm}

I am grateful to J.~Speth for the hospitality
at FZJ, J\"ulich, where this comment has been written.


\end{document}